\documentclass[%
 reprint,
superscriptaddress,
 amsmath,amssymb,
aip,
jcp,
]{revtex4-1}

\usepackage{graphicx}
\usepackage{dcolumn}
\usepackage{bm}
\usepackage{array}

\newcommand{\ket}[1]{\vert{#1}\rangle}

\begin{document}

\preprint{AIP/123-QED}

\title{Development of a general time-dependent absorbing potential for the CATM}
\author{Arnaud Leclerc}
\email{Arnaud.Leclerc@utinam.cnrs.fr}
\author{Georges Jolicard}
\affiliation{Institut UTINAM (CNRS UMR 6213, Universit\'e de Franche-Comt\'e, Observatoire de Besan\c con),\\
41bis Avenue de l'Observatoire, BP1615, 25010 Besan\c con cedex, France.}
\author{John P. Killingbeck}
\affiliation{Centre for Mathematics, University of Hull, Hull HU6 7RX, UK.}

\begin{abstract}
The Constrained Adiabatic Trajectory Method (CATM) allows us to compute solutions of the time-dependent Schr\"odinger
equation using the Floquet formalism and Fourier decomposition, using matrix manipulation within a non-orthogonal basis set, provided that suitable constraints can be applied to the initial conditions for the Floquet eigenstate. A general form is derived for the inherent absorbing potential, which can reproduce any dispersed boundary conditions. This new artificial potential acting over an additional time interval transforms any wavefunction into a desired state, with an error involving exponentially decreasing factors. Thus a CATM propagation can be separated into several steps to limit the size of the required Fourier basis. This approach is illustrated by some calculations for the $H_2^+$ molecular ion illuminated by a laser pulse.
\end{abstract}
\maketitle
\section{Introduction}

The development of new tools for solving quantum dynamical problems remains a very active field. One relatively modern method, the Constrained Adiabatic Trajectory Method (CATM), involves a Floquet processing \cite{shirley,chu,reviewguerin}
of the time-dependent Schr\"odinger equation (TDSE), using some form of artificial absorbing potential to connect a single Floquet eigenstate to the wavefunction of the system (namely the constrained Floquet state).
The Floquet formalism is frequently used to describe interactions between molecules and fields because it makes it possible to separate the quasi-periodic oscillations of the electromagnetic field from the slow, adiabatic variations of the envelope, allowing an efficient description of singular phenomena such as exceptional points \cite{Atabek201051}. In this paper we choose a slightly different approach, involving a rigorous periodicity, because all the time interval to be treated is included in the fundamental period, as in the $(t,t')$ approach
\cite{peskin}. The dynamical problem is thus transformed into a ``static'' problem, in the sense that the time will be included in an extended Hilbert space which is the product of the usual Hilbert space with the space of $T$-periodic functions.

This approach is different from the traditional one, in which the Hamiltonian is considered as constant over small time steps, with propagation of the wavefunction using a sequence of differential approximations. In the CATM the dynamical integration is completely replaced by the search for one eigenvector of a large matrix. That can be a difficult task, but many methods exist to solve this problem. We may use in particular the time-dependent wave operator theory \cite{reviewgeorges1,reviewgeorges2}, since the absorbing potential dilates the Floquet spectrum, isolating the required Floquet eigenvalue in a favourable manner \cite{CATM2}.

We briefly summarize the main ideas  of the CATM, already detailed in two articles\cite{CATM,CATM2}.
If the Hamiltonian of the unperturbed system is $H_0(q)$, with eigenvalues and eigenstates $\{E_j\;,|j\rangle\}$, and $W(q,t)$ is the time dependent part of the Hamiltonian, then we can work with the Floquet Hamiltonian in the extended Hilbert space:
\begin{equation}
 H_F(q,t)=H_0(q)+W(q,t)-i\hbar \frac{\partial}{\partial t}+{\cal V}(t).
\label{hamiltonian}
\end{equation}
Indeed, eigenvectors of $H_F$ are linked directly to solutions of the TDSE.
The fundamental period $T$ is here chosen as the entire duration $[0,T_0]$ of the physical interaction $W(t)$ plus an
artificial time interval $[T_0,T]$ ($T=T_0+\Delta T$), on which the absorbing potential ${\cal V}(t)$ will act.

The Floquet states are indexed with a double index \cite{reviewgeorges2} linked to the
molecular eigenstates (from $H_0$, $j\leftrightarrow \vert j \rangle$) and to an FBR Fourier basis set ($n \leftrightarrow \langle t\vert n\rangle = e^{-i n 2 \pi t/T}$). These eigenstates $\{ \vert \lambda_{j,n} (q,t) \rangle\}$ are defined by :
\begin{equation}
 H_F \vert  \lambda_{j,n} (q,t) \rangle  = E_{\lambda_{j,n}}  \vert \lambda_{j,n} (q,t) \rangle.
\label{eq2}
\end{equation}
If one of these eigenstates projected at $t=0$ is proportional to the selected initial value of the wavefunction $\Psi(0)$, then the knowledge of this eigenvector implies automatically that of the wavefunction driven by the Schr\"odinger equation. This follows from the rigorous expansion~\cite{reviewgeorges2}:
\begin{eqnarray}
\vert \Psi (q,t)\rangle &=& \sum_{j}   \langle \lambda_{j,n=0} (q,t=0)  \vert \Psi (q,t=0) \rangle  \nonumber \\
 & &\times \;  e^{-i E_{\lambda_{j,n=0}} t / \hbar } \vert \lambda_{j,0} (q,t) \rangle 
\end{eqnarray}
which can in this case be reduced to one single term:
\begin{equation}
 \vert \Psi(q,t) \rangle = \alpha \; e^{-i E_{\lambda} t /\hbar } \vert \lambda (q,t) \rangle
\label{wavefunction}
\end{equation}
with
\begin{equation}
\ket{\Psi(t=0)} = \alpha \langle t=0 \ket{\lambda} 
\label{condition}
\end{equation}
where the overlap $\alpha= \langle \lambda (t=0) \vert \Psi (t=0) \rangle$ is a complex number.
The time-dependent absorbing potential added to the real Hamiltonian ensures the approximate validity of Eq.\eqref{condition}.

In previous articles, two different forms for the time-dependent absorbing potential were proposed.
With this artificial potential the CATM was able to deal with many simple examples.
For a 2-level system, a non-diagonal matrix able to reproduce any initial condition has already been presented\cite{CATM2}.
Except for this particular case, until now the method could only integrate the dynamical evolution of a wavefunction initially projected onto one channel $j=l$ of the non-perturbed system $\ket{\Psi(0)}=\ket{l}$, because the absorbing potential was constructed to absorb all channels except the initial one. It was represented by a diagonal complex matrix, acting over the additional time interval $[T_0,T]$ and taking the form
\begin{equation}
 {\cal{V} }(t) = \sum_{j\neq l} V^c_{opt}(t) \vert j \rangle \langle j \vert \text{,}
\label{potentiel}
\end{equation}
with the three implicit conditions that 
$V^c_{opt}(t)$ is strictly null over $[0,T_0]$, $\Im(V^c_{opt}(t))$ is negative over $[T_0,T]$
and $ \vert \int_{T_0}^{T} \Im(V^c_{opt}(t)) dt \vert \gg \vert \Im (E_{\lambda}) \vert (T-T_0)$.
This form of potential allows us to propagate an initial condition $\ket{\Psi(0)}=\ket{l}$ since, owing to the periodic properties of Floquet eigenvectors, the final condition $\ket{\lambda(T_0+\Delta T)}\propto \ket{l}$, obtained with the introduction of $\mathcal{V}(t)$, induces the initial one $\ket{\lambda(0)}\propto \ket{l}$.

The subject of this paper is to find a general formula for an absorbing potential able to constrain
the Floquet eigenvector to any required boundary condition. 
With this improvement the CATM is able to treat dispersed initial wavefunctions and can also propagate the solution over very long time intervals (long laser pulse, train of pulses...), because the time interval can now be cut into smaller pieces $[0,T_1],[T_1,T_2],\dots$. 
So we need a new absorbing operator, acting in the extended Hilbert space and capable to transform any quantum state into a given different state, under the evolution dictated by the TDSE (for instance, able to transform $\ket{\Psi(t=T_2)}$ into $\ket{\Psi(T_2+\Delta T)}=\ket{\Psi(T_1)}$).

To this end, the paper is organized as follow.
In section \ref{derivation} we construct a treatment based on a non-orthogonal basis set, as a ``memory'' of the  initial state to be reproduced. Unfortunately this intermediate result is not exact and an improvement with a correction term is proposed in section \ref{rigorous}.
After that the asymptotic form of the Floquet eigenvector becomes conclusive. 
Some tests are made in section \ref{tests} on $H_2^+$ submitted to an intense laser pulse. For this same system we  compare two propagation schemes : a one-step scheme with the old version of the absorbing potential and a multi-step scheme with the new form of the absorbing potential.

\section{A heuristic definition of the absorbing potential \label{derivation}}

It is difficult to find directly an absorbing potential which acts selectively on each channel to impose the desired boundary conditions $\lambda(T_0+\Delta T)=\Psi(0)$. However, we can make profit from the previous definition of a ${\cal{V}}$ absorbing on all channels except one, by taking the initial wavefunction $\ket{\Psi(t=0)}$ in place of the state $\ket{l}$, at the expense of constructing a non-orthogonal basis set.

\subsection{Non-orthogonal basis set}

A non-orthogonal basis set for the Hilbert space ${\cal{H}}$ is defined as
\begin{equation}
 \vert \tilde{j} \rangle = 
\begin{cases}
 & \vert j \rangle \quad \forall j \neq l \\
 & \vert \Psi (0) \rangle \quad \text{ for } j=l
\end{cases}
\label{basenonorthogonale}
\end{equation}
where  $j$ is a global index for the free molecular eigenbasis.
$l$~is chosen among the $\{j\}$ so that $\vert \langle l \vert \Psi(0) \rangle \vert$ is a maximum, in order to
avoid numerical difficulties (such as divisions by too small numbers, because the later results will include $\langle l \vert \Psi(0)\rangle^{-1}$).
Thus the non-orthogonal basis is identical to the primitive orthogonal basis, except for one single
vector $\vert l\rangle$ which is replaced by the initial desired wavefunction $\vert \Psi(0)\rangle$.
In a sense, the initial wavefunction is ``kept in memory'' as one of the basis vectors by using this definition.
The new basis for the extended Hilbert space ${\cal{K}}={\cal{H}}\otimes{\cal{L}}_2(T)$ is thus:
\begin{equation}
 \vert \tilde{j} \rangle \otimes \vert t_i \rangle
\end{equation}
where $\vert t_i\rangle$ represents the DVR grid basis on time (the small $t_i$ being collocation points).
This can be written equivalently by Fourier transform on the FBR basis $\vert \tilde{j} \rangle \otimes \vert n \rangle$.

The transformation matrix $B$ from the orthogonal basis set $\{\vert k \rangle \otimes \ket{t_i}\}$ to the non-orthogonal one $\{\vert \tilde{j}\rangle \otimes \ket{t_{m}}\}$,
such that $\vert \tilde{j}\rangle \otimes \ket{t_i} = \sum_{km} B_{kj} \delta_{mi} \vert k \rangle \otimes \ket{t_m}$,
is constituted by a series of identical matrices, distributed along the main time diagonal $\delta_{mi}$.
Inside one block $t_i$ it is
\begin{equation}
B_{kj}=
\begin{cases}
  &\delta_{kj} \text{ if } j\neq l \\
  &\langle k \vert \Psi(0)\rangle \text{ if } j=l.
 \end{cases}
\label{passageB}
\end{equation}
For the inverse transformation matrix, one obtains:
\begin{equation}
(B^{-1})_{kj}=
\begin{cases}
  &\delta_{kj} \text{ if } j\neq l \\
  &-\frac{\langle k \vert \Psi(0)\rangle}{\langle l \vert \Psi(0)\rangle} \text{ if } j=l \text{ and } k\neq l \\
  & \frac{1}{\langle l \vert \Psi (0) \rangle} \text{ if } j=l \text{ and } k=l.
 \end{cases}
\label{passageBmoins1}
\end{equation}
These transformation matrices will be useful in the next sections.

\subsection{First conjecture}

Our first approach is to use the expression (\ref{potentiel}) without modification
within the non-orthogonal basis set,
so that an absorbing potential is placed on each channel $\vert \tilde{j}\rangle \neq\vert \tilde{l}\rangle$
but the channel $\vert \tilde{l}\rangle=\vert \Psi(0)\rangle$ is kept intact. 
We presume that this definition will produce the same effects as previously,
by giving an asymptotic absorption of components on all channels except the one which is
proportional to the desired initial wavefunction.
Such a potential expressed in the non-orthogonal basis of Eq.(\ref{basenonorthogonale}) is (see Tab.\ref{potbasenonorth}):
\begin{equation}
 {\cal{V}} = \sum_i\sum_{j\neq l} \vert \tilde{j}\rangle \vert t_i \rangle \langle t_i\vert\langle\tilde{j}\vert \;\;V^c_{opt}(t_i)
\label{premiereversion}
\end{equation}
where $V^c_{opt}(t)$ is different from zero only during the added artificial time interval $t\in[T_0,T]$.

\begin{table}[ht!]
\renewcommand{\arraystretch}{1.4}
\centering
{\scriptsize
\begin{center}
\begin{tabular}{|cccccc|}
\hline 1 & 0 & 0 & 0 & 0 & 0 \\
0 & $\ddots$ & 0 & 0 & 0 & 0  \\ 
0 & 0 & 1 & (col.$l$) & 0 & 0  \\ 
0 & (row $l$) & 0 & 0 & 0 & 0 \\ 
0 & 0 & 0 & 0 & 1 & 0 \\ 
0 & 0 & 0 & 0 & 0 & 1 \\
\hline
\end{tabular}
\end{center}
} 
\caption{Structure of the matrix which represents one block $t_i$ of the absorbing potential defined in Eq.(\ref{premiereversion})
within the non-orthogonal basis. This matrix multiplies the factor $V^c_{opt}(t_i)$ in Eq.(\ref{premiereversion}). All channels are expected to be
absorbed except the one ($l$) ``containing'' $\vert \Psi (0)\rangle$.}
\label{potbasenonorth}
\end{table}

\begin{table}[ht!]
\renewcommand{\arraystretch}{1.4}
\centering
{\scriptsize
\begin{center}
\begin{tabular}{|ccc|c|cc|}
\hline 
1 & 0 & 0 & (column $l$) & 0  & 0 \\
0 & $\ddots$ & 0 & $\vdots$ & 0 & 0 \\ 
0 & 0 & 1 & $-\frac{\langle j \vert \Psi (0) \rangle}{\langle l \vert \Psi(0) \rangle}$ & 0 & 0 \\ 
0 & 0 & 0 & 0 (row $l$) & 0 & 0 \\
0 & 0 & 0 & $-\frac{\langle j \vert \Psi (0) \rangle}{\langle l \vert \Psi(0) \rangle}$ & 1 & 0 \\
0 & 0 & 0 & $\vdots$ & 0 & 1 \\
\hline
\end{tabular}
\end{center}
} 
\caption{Structure of the matrix which represents one block $t_i$ of the absorbing potential in the orthogonal basis.
This matrix multiplies the factor $V^c_{opt}(t_i)$ in Eq.(\ref{eq61}).}
\label{nouveaupotentiel}
\end{table}

Further use of this non-orthogonal basis would create some complications in the Floquet Hamiltonian representation owing to the non-diagonal expression for $H_0$,
so we prefer to come back to the orthogonal basis for practical application.
The potential must thus undergo the following transformation:
\begin{equation}
 ({\cal{V}})^{orth.b.} = B\;\;({\cal{V}})^{non-orth.b.}\;\; B^{-1}.
\end{equation}
This is no longer a diagonal matrix, but now is, for each $t_i$ block (see Tab.\ref{nouveaupotentiel}):
\begin{eqnarray}
 \langle t_i \vert {\cal{V}} \vert t_i\rangle &=& V^c_{opt}(t_i) \times \label{eq61} \\
\sum_j \sum_k &&\left[ (1-\delta_{jl})\left( \delta_{jk} + \delta_{kl} \times \frac{-\langle j \vert \Psi(0)\rangle}
{\langle l \vert \Psi(0)\rangle}\right)\right] \vert j \rangle \langle k \vert. \nonumber
\end{eqnarray}

If this definition is appropriate the Floquet eigenvector calculated with the modified Hamiltonian
should satisfy the boundary conditions. In the most general form:
\begin{equation}
\langle j \vert \lambda (t=0)\rangle = \alpha \times \langle j\vert \Psi (0) \rangle \quad \forall \; j
\label{cigeneral}
\end{equation}
where $\alpha$ is a complex proportionality factor.

\subsection{Consequences of the first definition \label{section2c}} 

It is now necessary to analyse the effect of such a potential on Floquet eigenvectors, 
during the additional time.
We solve analytically the first order differential equation which drives the Floquet eigenvector components $\langle j \vert \lambda (t) \rangle$ over the time interval $[T_0,T]$ where the absorbing potential is different from zero, while simultaneously all couplings are reduced to zero. 
Remembering that Floquet eigenvectors satisfy, in the extended Hilbert space ${\cal{K}}$, the equation
\begin{equation}
 (H_F-E_{\lambda})\vert \lambda\rangle=0,
\end{equation}
projecting on the time and writing $\langle t\vert \lambda \rangle = \vert \lambda (t) \rangle$,
we get
\begin{equation}
 \dfrac{\partial}{\partial t} \vert \lambda (t) \rangle =
\frac{1}{i \hbar} \left[ H_0+{\cal{V}}(t)-E_{\lambda}\right]\vert\lambda (t)\rangle.
\label{eq13vp}
\end{equation}
For the time being, ${\cal{V}}$ is the operator defined in Eq.(\ref{premiereversion}).
Eq.\eqref{eq13vp} with $\mathcal{V}(t)$ expressed in the orthogonal basis (see Eq.\eqref{eq61}) has an analytical integral solution involving $E_{\lambda}$ as an undetermined parameter (implicit solution).

Using the notations $\langle j \vert \lambda \rangle =  \lambda_j$ and $\langle j \vert \Psi (0) \rangle = \psi^0_j $,
one can write :

\begin{widetext}
\begin{equation}
\dfrac{\partial}{\partial t}\left(
\begin{array}{l}
 \lambda_1 (t) \\ 
 \lambda_2 (t) \\ 
\vdots \\ 
 \lambda_j (t) \\ 
\vdots
\end{array} \right)
= 
\dfrac{1}{i \hbar} \left[ \left(
\begin{array}{llll}
E_1 &  &  & 0 \\ 
 & \ddots &  &  \\ 
 &  &  E_j & \\ 
0 &  &  & \ddots
\end{array}
\right)
- E_{\lambda}  + V^c_{opt}(t)  \times  \left( 
\begin{array}{lllll}
1 &  & 0 & \text{(col.l)} &  \\ 
 & 1 &  & \frac{-\psi^0_j}{\psi^0_l} & 0 \\ 
 &  & \ddots & \vdots &  \\ 
 & 0 &  & 0 & \text{(lin.l)} \\ 
 &  &  & \frac{-\psi^0_j}{\psi^0_l} & 1
\end{array}
\right) \right] 
\left(
\begin{array}{l}
 \lambda_1 (t) \\ 
 \lambda_2 (t) \\ 
\vdots \\ 
 \lambda_j (t) \\ 
\vdots
\end{array} \right).
\end{equation}
\end{widetext}

Thus for all $t$ within $[T_0,T]$ we have
\begin{eqnarray}
 \frac{\partial}{\partial t} \lambda_j (t) &=& \frac{1}{i \hbar}(E_j-E_{\lambda}) \lambda_j (t)  \nonumber \\
& & + \frac{1}{i \hbar}
V^c_{opt}(t)\left[\lambda_j(t)- \frac{\psi^0_j}{\psi^0_l} \lambda_l (t)\right].
\label{eqdiffj}
\end{eqnarray}
In the particular case of the line $j=l$, this equation becomes simply :
\begin{equation}
 \frac{\partial}{\partial t} \lambda_l (t) = \frac{1}{i\hbar}(E_l-E_{\lambda}) \lambda_l (t)\text{,} 
\label{caspartl}
\end{equation}
which directly gives
\begin{equation}
\lambda (t)_l=\lambda_l (t=T_0) \exp\left(\frac{i}{\hbar}(E_{\lambda}-E_l)(t-T_0)\right) \text{, }
\label{caspartsol}
\end{equation}
while Eq.(\ref{eqdiffj}) for $j\neq l$ is somewhat more complicated. 
After some elementary simplifications one obtains a formula describing the Floquet eigenvectors behaviour when
$t \in [T_0,T]$ for $j\neq l$:
\begin{eqnarray}
\forall j \neq l \quad && \lambda_j (t) = \lambda_j (T_0) 
\times e^{\frac{i}{\hbar} \int_{T_0}^t (E_{\lambda}-E_j-V^c_{opt}(t'))dt'} \nonumber \\
+ \frac{i}{\hbar} \frac{\psi^0_j}{\psi^0_l}&& \lambda_l(T_0) \left[
\int_{T_0}^t V^c_{opt} (t') e^{\frac{i}{\hbar}\int_{T_0}^{t'} (E_j - E_l + V^c_{opt}(t''))dt''} dt'
\right] \nonumber \\
&& \times e ^{\frac{i}{\hbar} \int_{T_0}^t
(E_{\lambda}-E_j-V^c_{opt}(t'))dt'}. \label{casgensol}
\end{eqnarray}
(Note that Eq.(\ref{caspartsol}) can also be obtained as a particular case of Eq.(\ref{casgensol})).
At this point we can ask whether Eq.\eqref{casgensol} is consistent with the equality $\vert\lambda(t=T)\rangle = \alpha \;\vert \Psi(0) \rangle $.
One can first transform the expected proportionality relationships \eqref{condition} into an equality:
\begin{equation}
\ket{\tilde{\lambda}(t=0)}=\ket{\tilde{\lambda}(t=T)}=\ket{\Psi(0)}
\label{boundary}
\end{equation}
by introducing the vector
\begin{equation}
\vert \tilde{\lambda} (t)\rangle = 
 \vert \lambda(t) \rangle \times \frac{\psi_l^0}{\lambda_l(t)}
\end{equation}
with the components (see Eq.\eqref{casgensol})
\begin{equation}
\tilde{\lambda}_j (t) =
\begin{cases}
A(t) + \psi^0_j \times B(t) \text{ if } j\neq l\\
\psi^0_l \text{ if } j=l
\end{cases}
\end{equation}
and with
\begin{eqnarray}
A(t) &=& \frac{ \lambda_j(T_0) \psi^0_l}{\lambda_l(T_0)} 
e^{\frac{i}{\hbar} \int_{T_0}^t (E_l-E_j-V^c_{opt}(t'))dt'} \\
B(t) &= &\frac{i}{\hbar} \int_{T_0}^t V^c_{opt} (t') e^{\frac{i}{\hbar}\int_{T_0}^{t'}
(E_j-E_l+V^c_{opt}(t''))dt''} dt' \nonumber \\ 
&&\times e^{\frac{i}{\hbar}\int_{T_0}^t(E_l-E_j-V^c_{opt}(t'))dt'}.
\end{eqnarray}
Thus, the eigenvector will satisfy the expected boundary conditions \eqref{boundary} only if we have
\begin{equation}
\begin{cases}
A(t=T) \simeq 0 \\
B(t=T) \simeq 1.
\end{cases}
\end{equation}
The first condition implies that $\int_{T_0}^T\Im(E_l-E_j-V^c_{opt}(t'))dt'$ is large,
creating an strong exponential decrease.
However a detailed analysis of the second function $B$ remains difficult, because of the double interlocked integral.
Here it is impossible to indicate clearly when the second condition is respected.
However, numerical tests reveal that there is a non-perfect control of the boundary conditions.
It thus seems necessary to introduce a correction term to the definition of ${\mathcal{V}}(t)$ in order to enforce the required boundary conditions.

\section{A rigorous definition able to reproduce any initial condition \label{rigorous}}

To solve this problem rigorously, we follow an inverse solution. Considering the results of section \ref{section2c}, it is possible to change the solution $\vert \lambda (t) \rangle$ slightly 
to obtain a proportionality to the initial conditions,
and to find afterwards what should be added to the absorbing potential so as to obtain this modified result.
That is, we start from the solution to go back to the appropriate equation.

\subsection{Ideal solution}

If we add in Eq.\eqref{casgensol} a factor $(V^c_{opt}(t)+E_j-E_l)$ at the ``feet'' of the integral,
instead of $V^c_{opt}(t)$, i.e.
\begin{eqnarray}
&\int_{T_0}^t V^c_{opt} (t') &e^{\frac{i}{\hbar}\int_{T_0}^{t'} (E_j - E_l + V^c_{opt}(t''))dt''} dt'\label{hyp1}\\
&&\downarrow \nonumber \\
&\int_{T_0}^t (E_j - E_l& + V^c_{opt} (t')) e^{\frac{i}{\hbar}\int_{T_0}^{t'} (E_j - E_l + V^c_{opt}(t''))dt''} dt'\label{hyp2}
\end{eqnarray}
then this integral simplifies in the following way:
\begin{eqnarray}
&\int_{T_0}^t (V^c_{opt} (t') + E_j - E_l) e^{\frac{i}{\hbar}\int_{T_0}^{t'} (E_j - E_l + V^c_{opt}(t''))dt''} dt'& \nonumber\\
&=  \int_{T_0}^t \frac{\hbar}{i} \frac{\partial}{\partial t'} e^{\frac{i}{\hbar}\int_{T_0}^{t'} (E_j - E_l + V^c_{opt}(t''))dt''} dt'& \nonumber\\
&= \frac{\hbar}{i} \left( e^{\frac{i}{\hbar}\int_{T_0}^{t} (E_j - E_l + V^c_{opt}(t'))dt'} - 1 \right) \label{simpint} &
\end{eqnarray}
and the solution for $ \lambda_j (t)$ takes the form:
\begin{eqnarray}
 \lambda_j (t) &=&
 \lambda_j (T_0) e^{\frac{i}{\hbar} \int_{T_0}^t (E_{\lambda} - E_j - V^c_{opt}(t')) dt'}  \label{eq42}\\
+ \lambda_l (T_0) &\frac{\psi^0_j}{\psi^0_l}& \left(
e^{\frac{i}{\hbar} ( E_{\lambda}-E_l ) (t-T_0)} - e^{\frac{i}{\hbar} \int_{T_0}^t (E_{\lambda} - E_j - V^c_{opt}(t')) dt'} \right) \nonumber
\label{conjecture}
\end{eqnarray}
As in the previous section, we introduce
\begin{equation}
\vert \tilde{\lambda} (t)\rangle = 
 \vert \lambda(t) \rangle \times \frac{\psi_l^0}{\lambda_l(t)}
\end{equation}
with components
\begin{equation}
\tilde{\lambda}_j (t) =
\begin{cases}
C(t) + \psi^0_j \times D(t) \text{ if } j\neq l\\
\psi^0_l \text{ if } j=l
\end{cases}
\end{equation}
where
\begin{eqnarray}
C(t) &=& \frac{ \lambda_j(T_0) \psi^0_l}{\lambda_l(T_0)} 
e^{\frac{i}{\hbar} \int_{T_0}^t (E_l-E_j-V^c_{opt}(t'))dt'} \\
D(t) &= &1- e^{\frac{i}{\hbar}\int_{T_0}^t(E_l-E_j-V^c_{opt}(t'))dt'}\text{.}
\end{eqnarray}
The asymptotic condition $\ket{\tilde{\lambda}(T)}=\ket{\Psi(0)}$ becomes easier to satisfy: we require
\begin{equation}
\begin{cases}
C(t=T) \simeq 0 \\
D(t=T) \simeq 1
\end{cases}
\end{equation}
this is equivalent to
\begin{subequations}
\begin{eqnarray}
 e^{\frac{i}{\hbar}\int_{T_0}^T(E_l-E_j-V^c_{opt}(t'))dt'}\rightarrow 0 \\
\Leftrightarrow \, \int_{T_0}^{T}\Im(E_l-E_j-V^c_{opt}(t))dt \gg 1 \text{.} \label{integrale}
\end{eqnarray}
\end{subequations}
At this stage, these corrections (Eqs. \ref{hyp1}-\ref{hyp2}) are still inconsistent with the selected absorbing potential.
We see now what change we should make in the absorbing potential so as to produce a favourable conclusion after integration of the differential equation \eqref{eq13vp}.

\subsection{A correction term to the first definition}

Progressively tracing back the solution, we deduce that all we have to do is to introduce
a corrective energy term $(E_j-E_l)$ inside the absorbing potential. The matrix representation is shown on Tab. \ref{derniere}
for each Floquet block $t_i$. The definition \eqref{eq61} is thus replaced by
\begin{widetext}
\begin{equation}
 \langle t_i \vert {\cal{V}} \vert t_i\rangle = \sum_j \sum_k 
\left[ (1-\delta_{jl})\left( \delta_{jk} V^c_{opt}(t_i) + \delta_{kl} \times 
\frac{-\langle j \vert \Psi(0)\rangle}{\langle l \vert \Psi(0) \rangle} \left( V^c_{opt}(t_i)+E_j-E_l \right) \right)\right] \vert j \rangle \langle k \vert.
\label{bonneformule}
\end{equation}
\end{widetext}

\begin{table}[htp]
\centering
{\scriptsize
\begin{center}
\renewcommand{\arraystretch}{2}
\begin{tabular}{|ccc|c|c|}
\hline $V^c_{opt}(t_i)$ & 0 & 0 & (column $l$) & 0 \\
0 & $V^c_{opt}(t_i)$ & 0 & $-\frac{\langle j \vert \Psi (0) \rangle}{\langle l \vert \Psi(0) \rangle} \times 
\left( V^c_{opt}(t_i) + E_j - E_l \right)$ & 0 \\ 
0 & 0 & $\ddots$ & $\vdots$ & 0 \\ 
0 & 0 & 0 & 0 (row $l$) & 0 \\ 
0 & 0 & 0 & $-\frac{\langle j \vert \Psi (0) \rangle}{\langle l \vert \Psi(0) \rangle}\times 
\left( V^c_{opt}(t_i) + E_j - E_l \right)$ & $V^c_{opt}(t_i)$ \\
\hline
\end{tabular}
\end{center}
} 
\caption{Matrix representation of one block $t_i$ of the absorbing potential (second version)
within the orthogonal basis set.}
\label{derniere}
\end{table}

Eq.(\ref{eq13vp}) for the Floquet vector components is replaced by
\begin{eqnarray}
 \frac{\partial}{\partial t} \lambda_j (t) &=& \frac{1}{i \hbar}(E_j-E_{\lambda}+ V^c_{opt}(t) ) \lambda_j (t)  \nonumber\\
&+ \frac{1}{i \hbar} &\left[ - \frac{\psi^0_j}{\psi^0_l} (V^c_{opt}(t) + E_j -E_l) \right] \lambda_l (t).
\label{eqdiffjderniere}
\end{eqnarray}
After integration, the formulae can be simplified, as indicated in Eq.(\ref{simpint}).
Finally we obtain the ideal solution of Eq.(\ref{cigeneral}), solving the differential equations \eqref{eqdiffjderniere} in presence of the correcting term and taking $t=T$:
\begin{eqnarray}
 \lambda_j (T) &=&
 \lambda_j (T_0) e^{\frac{i}{\hbar} \int_{T_0}^T (E_{\lambda} - E_j - V^c_{opt}(t')) dt'}  \label{eq42}\\
+ \lambda_l (T_0) &\frac{\psi^0_j}{\psi^0_l}& \left(
e^{\frac{i}{\hbar} ( E_{\lambda}-E_l ) (T-T_0)} - e^{\frac{i}{\hbar} \int_{T_0}^T (E_{\lambda} - E_j - V^c_{opt}(t')) dt'} \right) \nonumber
\end{eqnarray}
The desired asymptotic behaviour
\begin{eqnarray}
\langle j \vert \lambda (t=0) \rangle &=& \alpha \times \psi^0_j \nonumber\\
\text{with } \alpha &=&\frac{\lambda_l (T_0)}
{\psi^0_l} e^{\frac{i}{\hbar} (E_ {\lambda}-E_l)(T-T_0)}
\end{eqnarray}
is obtained if the factor $e^{\frac{i}{\hbar} \int_{T_0}^T (E_l - E_j - V^c_{opt}(t')) dt'}$ tends properly to zero. It means that the integrated surface of $V^c_{opt}(t)$ over the time dimension must possess a large negative imaginary part, so that a  strongly decreasing real exponential appears, becoming almost zero when $t=T$:
\begin{equation}
e^{\frac{i}{\hbar} \int_{T_0}^T \Re (E_l - E_j - V^c_{opt}(t') ) dt'} e^{-\frac{1}{\hbar} \int_{T_0}^T \Im (E_l - E_j - V^c_{opt}(t') ) dt'} \simeq 0.
\label{ddet}
\end{equation}
In this formulation the CATM will also run with any initial superposition of states.

\section{Tests on $H_2^+$ submitted to an intense laser pulse \label{tests}}

\subsection{Illustration of the integration scheme}

\begin{figure}[!ht]
 \centering
\includegraphics[width=\linewidth]{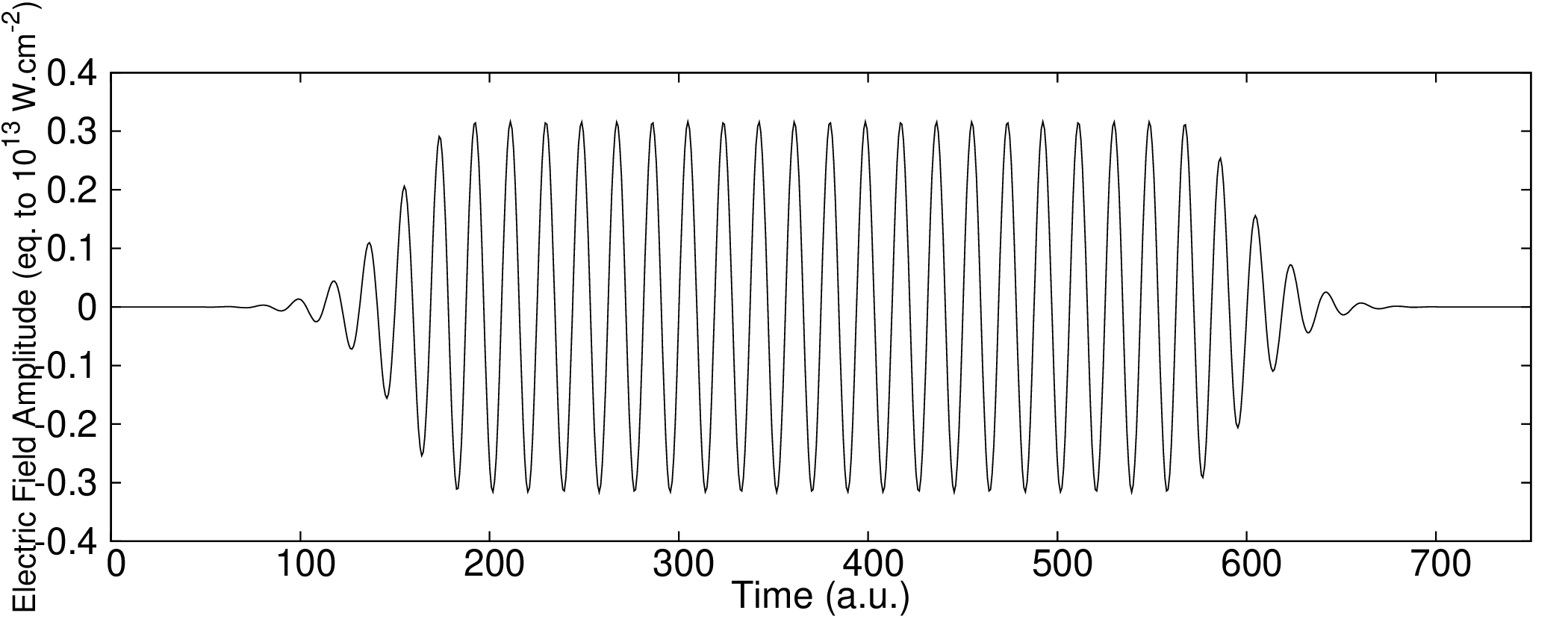}
 \caption{Pulse with a total duration of $750$a.u., carrier wave frequency $0.335$a.u. and intensity $10^{13}W.cm^{-2}$.}
 \label{champ}
\end{figure}

Using this new form for the absorbing potential, we have deal with a test-system.
Many propagations were made on the example of the $H_2^+$ molecular ion, modelled by
its first two electronic surfaces $^2\Sigma_g^+$ and $^2\Sigma_u^+$ and illuminated by
an intense laser pulse of total duration $750$a.u. ($18$fs) with a carrier wave frequency $0.335$a.u. (i.e. a wavelenght of $136$nm) and an intensity of $10^{13}W.cm^{-2}$.
This pulse is represented in Fig.\ref{champ}.
The initial vibrational state was chosen as $v=2$ (second excited state) because beginning
in this state with this precise carrier frequency and this relatively high intensity leads to an electromagnetic trapping \cite{theseviennot}. This choice makes the test more significant since the CATM algorithm should recur strongly non-linear phenomena. 

\begin{figure}[!ht]
 \centering
\includegraphics[width=\linewidth]{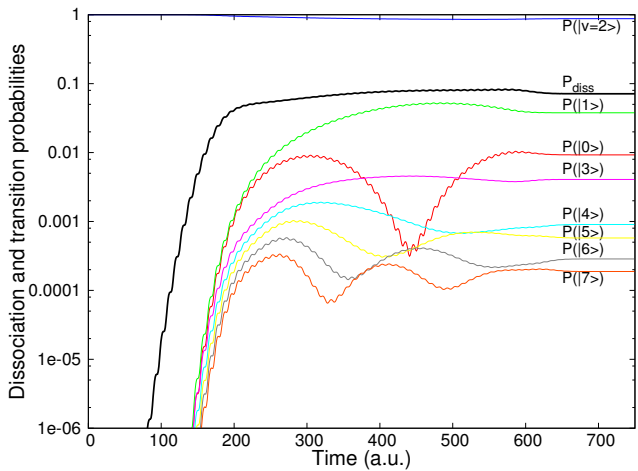}
 \caption{Dissociation (black) and transition (colours) probabilities $P(\ket{v})=|\langle v \ket{ \Psi (t)}|^2$ for $H_2^+$ submitted to the pulse of Fig.\ref{champ}. 
The initial state was the second exited state $v=2$.}
 \label{enfin}
\end{figure}

Indeed, Fig.\ref{enfin} shows that in these intense pulse condition the dissociation remains small (final $P_{diss}\simeq 7.14\times10^{-2}$).
The initial state keeps the main part of the population, while states $1$ reaches $5\%$ at its maximum. The fundamental state population shows important oscillations between $5\times10^{-4}$ and $10^{-2}$. Other bound states, with higher energies than the initial state, show many oscillations in population, with a frequency which increases gradually as one moves up in energy.

Since our principal aim is to test the CATM method in its more general formulation, we divide the time interval
into $N_s$ steps and we realize a CATM propagation for each step $[0,T_1],[T_1,T_2],[T_2,T_3]\dots$ using the general formula of Eq.(\ref{bonneformule}). 
Each interval $[T_i,T_{i+1}]$ is successively considered as the physical interval to be submitted to the CATM procedure, which requires for each step the addition of an artificial time interval of duration $\Delta T$, on which the absorbing potential is introduced.
We just have to make the correspondance $t'=t-T_i$ (i.e. $T_i\leftrightarrow0$, $T_{i+1}\leftrightarrow T_0$ in all formulae in section \ref{rigorous}).

The first step is a propagation issuing from a single eigenstate $\ket{\Psi(0)}=\ket{v=2}$, during which the wavefunction is dispersed over all the basis set. Then for the following steps, the CATM must ensure the correct reconstruction of the totally dispersed initial states ($\ket{\Psi(T_i)}=\sum_v c^i_v \ket{v}$). These more general problems are rigorously solved provided that Eq\eqref{integrale} is satisfied. This is the case if we use the modified expression \eqref{bonneformule} for the absorbing potential; however the supplementary terms $(E_j-E_l)$ that it includes create numerical difficulties.

For each step, over the additional time interval $\Delta T$, terms
$\frac{-\langle j \vert \Psi(0)\rangle}{\langle l \vert \Psi(0) \rangle} \left(E_j-E_l \right)$
are present in the column number $l$ of the Hamiltonian, where $\Psi(0)$ is the wave function at the initial instant of the step. Given that the propagation is made using numerous Fast Fourier Transforms (FFT), these terms must
be multiplied by an Heaviside function of time which is zero everywhere except on the additional interval 
of duration $\Delta T$.
We thus encounter some numerical difficulties because the Heaviside function is discontinuous and the FFT needs very high basis frequencies to describe correctly such discontinuities (Gibbs phenomenon).
Moreover, the electric field must be zero over the artificial time interval, so as to avoid transitions during the absorption phase.
To limit this problem we choose the time step separation carefully, corresponding to nodes of the electric field.
This is the easy way but it does not restrict the generality of the approach. In principle, it would be also possible to choose other step positions with non-zero values of the electric field, but in this case we would have to ensure the continuity with an artificial return to zero during the additional time interval.

In this framework, the determination of one Floquet eigenvector was realized with a wave operator algorithm \cite{reviewgeorges2},
starting from a test vector which is constant over the time step and  modifying it iteratively with a RDWA (Recursive Distorted Wave Approximation) procedure.

\begin{figure}[!ht]
 \centering
\includegraphics[width=\linewidth]{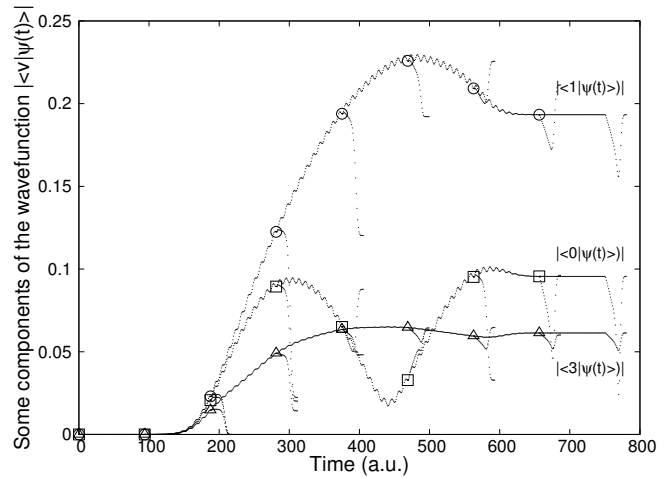}
 \caption{Evolution of some components of the wavefunction on bound states of $H_2^+$, $|\langle v \ket{\Psi (t)}|$, $v=0,1,3$. 
Illustration of the CATM scheme with a decomposition of the propagation into 8 big steps, $\simeq94a.u.$ are treated for each CATM step. Each point in the figure corresponds to a grid point of the time-discretization.
Grid points obtained during additional time intervals are kept here to help in understanding the diagram.}
 \label{plusieurspas}
\end{figure}

\begin{figure}[!ht]
 \centering
\includegraphics[width=\linewidth]{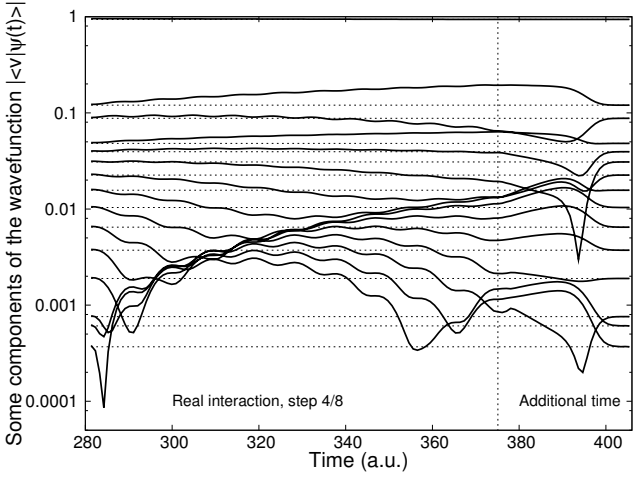}
 \caption{``Zoom in'' on the fourth step. Evolution of some components of the wavefunction $|\langle v \ket{\Psi (t)} |$, $v=0\dots14$ (same example as in the previous figure). The real interaction is extended with an artificial interval over which the required initial conditions are progressively recovered.}
 \label{zoom}
\end{figure}

To illustrate the principles of the method, Fig. \ref{plusieurspas} shows results for a CATM calculation made with an 8-step propagation.
At each step the absorbing potential constrains the wavefunction components to match the values obtained
at the end of the previous step. The small supplementary pieces of curve correspond to additional
time intervals during which the absorbing potential is present.
Fig.\ref{zoom} is a ``zoom in'' on these results for one given step.

\subsection{The stability of multi-step propagations}

We now compare three different calculations made using the following parameters:
run A, one step with $N=1024$ Fourier basis functions;
run B, two steps with $N=512$ and 
run C, four steps with $N=256$.
For each calculation the absorbing potential amplitude was chosen sufficiently large
to correctly reproduce the initial conditions, so that the jumps between
each step and the differences between the A, B, C curves become both stable and negligible.
In other words, the time integral of the absorbing potential was always sufficient to 
stabilize the numerical values of the calculated probabilities in the three different configurations. 

\begin{figure}[!ht]
 \centering
\includegraphics[width=\linewidth]{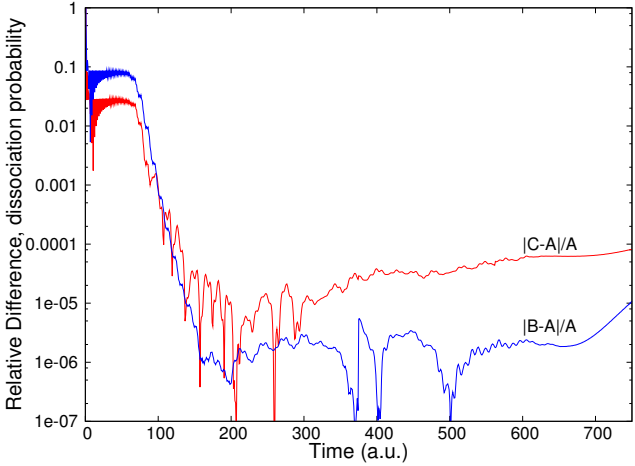}
 \caption{Relative difference of dissociation probability for $H_2^+$ vs. time, between runs A (1 step), B (2 steps) and C (3 steps),
 $\vert \frac{B-A}{A} \vert$ (blue) and $\vert \frac{C-A}{A} \vert$ (red).}
 \label{ecartpdiss}
\end{figure}

Previous calculations made on this system using a one step procedure \cite{CATM} have proved that the CATM results are very precise when compared with ones obtained by using standard wave-packet propagation techniques.
Thus run A is chosen as the reference and relative differences are successively computed between A and B and between A and C. Their evolutions are shown in Fig.\ref{ecartpdiss}. Since the time when the dissociation probability is no longer
negligible ($t\simeq125$a.u.), differences become rapidly small and stay so. We cannot see any significant increasing
at the moment of step changes ($t=187.5$a.u., $375$a.u. et $562.5$a.u.),
although a small discontinuity is present at $t=375$a.u. for the run B.

\begin{figure}[!ht]
 \centering
\includegraphics[width=\linewidth]{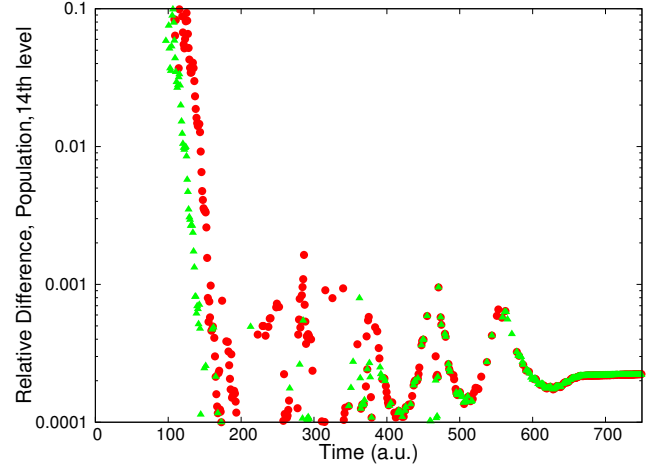}
 \caption{Evolution of the relative difference on the transition probability $\vert \langle v=13 \vert \Psi(t) \rangle \vert$, between runs A (1 step), B (2 steps) and C (3 steps),
 $\vert \frac{B-A}{A} \vert$ (green triangles) and  $\vert \frac{C-A}{A} \vert$ (red points). 
}
 \label{ecartptrans}
\end{figure}

One example of a relative difference, that for the transition probability to state $v=13$, is
presented in Fig.\ref{ecartptrans}. It varies inversely with the probability itself
and it is always lower than $0.2\%$ (except for negligible values of probability), despite the very small value of the final probability ($4.47\times10^{-6}$).

\begin{figure}[!ht]
 \centering
\includegraphics[width=\linewidth]{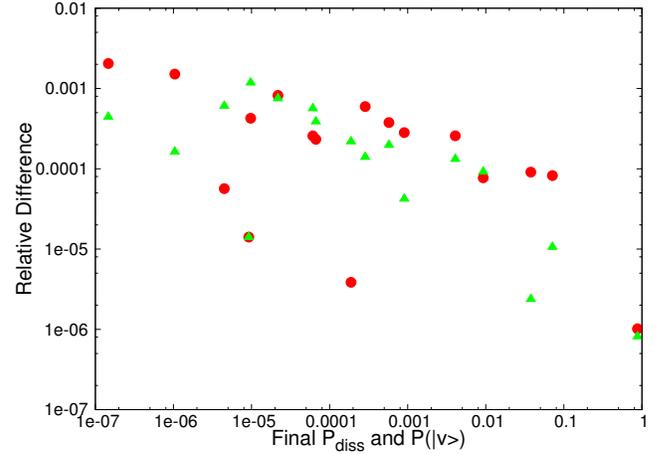}
 \caption{Comparison of the relative difference on final dissociation probability (right bottom) $|P_{\text{diss}}^{B,C}-P_{\text{diss}}^A|/P_{\text{diss}}^A$, for $t=750$a.u. and relative difference on transition probabilities to the 16 first bound states (other points) $|P^{B,C}(\ket{v})-P^A(\ket{v})|/P^A(\ket{v})$, between runs A (1 step), B (2 steps) and C (3 steps),
 $\vert \frac{B-A}{A} \vert$ (green triangles) and  $\vert \frac{C-A}{A} \vert$ (red points).}
 \label{ecartsfinaux}
\end{figure}

Finally Fig.\ref{ecartsfinaux} shows a comparison of the final relative differences ($t=750$a.u.)
of dissociation and transition probabilities obtained with runs A, B and C.
The precision is stable in relation to scale changes and it stays lower than $0,2\%$ 
once the probability in question is greater than $10^{-7}$.
Relative differences at very small probabilities are almost of the same magnitude as those at large probabilities.

\section{conclusion}

We find that it is possible to drive the CATM with multi-step propagation, which allows us to decrease the dimension of the Fourier basis used to describe the time variations. The multi-step propagation with the CATM is made possible by the development of a general form for the time-dependent absorbing potential, which can now constrain the Floquet state connected to any general initial superposition of states. The operating condition remains the same as for the simple case of an initial eigenstate of the free system, i.e. the time integral of the imaginary part of the absorbing potential must be negative and large enough. The multi-step CATM preserves the two principal features of the one-step version: contrary to the standard wave packet propagations, it calculates the probabilities with an almost constant relative accuracy, whether large or small and it produces a dilatation of the $H_F$-spectrum in the complex plane and thus facilitates recursive treatment by the RDWA of wave operator theory.
However these favourable results should not make us believe that the multi-step process always behaves perfectly. Indeed, sometimes the Gibbs phenomenon prevented us from easily obtaining a very good convergence.
We are currently working to circumvent these difficulties due to discontinuous functions which limit
the convergence speed and  radius. 
One possible approach would involve slowly collapsing the physical hamiltonian to zero before the beginning of the absorption, and even the diagonal terms during the artificial extra time. This idea is currently being tested.
The CATM should also soon be adapted to a spatial description on a DVR grid.

Despite computational difficulties, our promising results prompt us to continue the exploration of the method,
both to develop a more general theory and to make further applications.
In principle the absorbing potential presented here is efficient enough to constrain any statevector to be transformed
progressively into a given different one, as driven by the TDSE. This was the necessary condition to be able to propagate such an initial state with the CATM. The technical advantage of this is that it transforms a dynamical integration problem into an eigenvalue problem.

%

\end{document}